\documentclass[prd,aps,showpacs,tightenlines,nofootinbib,preprint,preprintnumbers]{revtex4}
\usepackage{graphicx}
\usepackage{amsfonts}
\usepackage{amssymb}
\usepackage{latexsym}
\usepackage{cancel}
\usepackage{color}

 \newcommand{\CL}{{\cal L}}
 
\newcommand{\CO}{{\cal O}}

\newcommand{\bear}{\begin{array}}  \newcommand{\eear}{\end{array}}
\newcommand{\bea}{\begin{eqnarray}}  \newcommand{\eea}{\end{eqnarray}}
\newcommand{\beq}{\begin{equation}}  \newcommand{\eeq}{\end{equation}}
\newcommand{\bef}{\begin{figure}}  \newcommand{\eef}{\end{figure}}
\newcommand{\bec}{\begin{center}}  \newcommand{\eec}{\end{center}}
\newcommand{\non}{\nonumber}  
\newcommand{\lmk}{\left(}  \newcommand{\rmk}{\right)}
\newcommand{\lkk}{\left[}  \newcommand{\rkk}{\right]}
  
\newcommand{\del}{\partial}  
\newcommand{\vect}[1]{\mbox{\boldmath${#1}$}}
\newcommand{\bib}{\bibitem} 
\newcommand{\la}{\left\langle} \newcommand{\ra}{\right\rangle}

\def\IB#1#2#3{{\bf #1}, #2 (19#3)}
\def\IBB#1#2#3{{\bf #1}, #2 (20#3)}
\def\IBID#1#2#3{{\it ibid}. {\bf #1}, #2 (19#3)}
\def\IBIDD#1#2#3{{\it ibid}. {\bf #1}, #2 (20#3)}

\def\JETP#1#2#3{JETP. {\bf #1}, #2 (19#3)}

\def\JHEPP#1#2#3{J. High Energy Phys. {\bf #1}, #2 (20#3)}

\def\JP#1#2#3{J. Phys. A {\bf #1}, #2 (19#3)}

\def\MNRASS#1#2#3{Mon. Not. R. Astron. Soc. {\bf #1}, #2 (20#3)}

\def\NPB#1#2#3{Nucl. Phys. {\bf B#1}, #2 (19#3)}

\def\PLA#1#2#3{Phys. Lett. A {\bf #1}, #2 (19#3)}
\def\PLB#1#2#3{Phys. Lett. B {\bf #1}, #2 (19#3)}
\def\PLBB#1#2#3{Phys. Lett. B {\bf #1}, #2 (20#3)}
\def\PLBold#1#2#3{Phys. Lett. {\bf#1B}, #2 (19#3)}

\def\PRD#1#2#3{Phys. Rev. D {\bf #1}, #2 (19#3)}
\def\PRDD#1#2#3{Phys. Rev. D {\bf #1}, #2 (20#3)}

\def\PRL#1#2#3{Phys. Rev. Lett. {\bf#1}, #2 (19#3)}

\def\PTP#1#2#3{Prog. Theor. Phys. {\bf #1}, #2 (19#3)}

\def\RPP#1#2#3{Rep. Prog. Phys. {\bf #1}, #2 (19#3)}

\begin{document}

\title{Cosmological evolution of cosmic strings with time-dependent
tension}
\author{Masahide Yamaguchi}
\affiliation{Department of Physics and Mathematics, Aoyama Gakuin
University, Sagamihara 229-8558, Japan}

\date{\today}


\begin{abstract}
We discuss the cosmological evolution of cosmic strings with
time-dependent tension. We show that, in the case that the tension
changes as a power of time, the cosmic string network obeys the scaling
solution: the characteristic scale of the string network grows with the
time. But due to the time dependence of the tension, the ratio of the
energy density of infinite strings to that of the background universe is
{\it not} necessarily constant.
\end{abstract}


\maketitle


\section{Introduction}

Cosmic strings are produced in field theories, through various thermal
\cite{Kibble} or nonthermal \cite{KVY} phase transitions, as a result of
spontaneous symmetry breaking. Recently, cosmic superstrings have also
received much attention, partly because they are produced after brane
inflation \cite{css}.

Cosmic strings have fruitful implications for cosmology. They can
generate density fluctuations, gravitational waves, gravitational
lensing effects, and so on (for review, see \cite{VS,HK}). Particularly,
they were investigated as possible seeds of galaxies and large scale
structure. Unfortunately, recent observations of the cosmic microwave
background radiation (CMBR) anisotropy by the Wilkinson Microwave
Anisotropy Probe (WMAP) revealed that cosmic strings cannot become a
dominant source of primordial density fluctuations \cite{WMAP}. Only a
$~10$\% contribution by cosmic strings to the CMBR angular power
spectrum is allowed on large scales \cite{AR,Pogosian,Endo}. However,
they can still play an important role in large scale structure formation
because the density fluctuations generated by cosmic strings are
peculiar, that is, incoherent, nonadiabatic, and highly non-Gaussian. In
fact, cosmic strings may cause the early star formation and early
reionization observed by WMAP \cite{AR,ER}. Furthermore, the
contribution from cosmic strings to the galaxy power spectrum may loosen
the bounds on the sum of the neutrino masses, and this possibility is
discussed \cite{BMY}.

The cosmological evolution of cosmic strings has been studied for a long
time. The key property is scaling, when the typical length of the cosmic
string network grows with the horizon scale. Then, the number of
infinite strings per horizon volume is a constant irrespective of time
and hence the ratio of the energy density of infinite strings to that of
the background universe is a constant. Such a scaling solution is
confirmed both analytically \cite{Kibble} and numerically
\cite{AT,BB,AS} by use of the Nambu-Goto action \cite{NG}, which is an
effective action obtained after integrating out heavy modes of particles
and neglecting high curvature of the geometry.\footnote{This scaling
property is confirmed not only for cosmic (local) strings but also for
global strings \cite{gs} and global monopoles \cite{gm}.} Based on this
property, many cosmological implications have been discussed, as stated
above.

In the past the constancy of the tension of cosmic strings has been
assumed in order to derive the scaling solution. However, in
cosmological situations the tension of the strings can depend on the
cosmic time. Let's consider a complex scalar field $\phi$, whose
effective potential depends on its interactions and can be described by
\beq
  V(\phi) = \frac{\lambda}{4} (|\phi|^2 - \eta^2)^2,
  \label{eq:potential}
\eeq
where $\eta$ is the effective expectation value of $|\phi|$. Then, the
typical radius of a string $\delta_s$ is given by $\delta_s \simeq
1/(\sqrt{\lambda} \eta)$ and the tension $\mu$ is given by $\mu \simeq
\eta^2$. Note that the tension $\mu$ does not depend on the coupling
constant but on the expectation value alone. For example, assume that
$\phi$ has a wine-bottle potential at zero temperature. The finite
temperature effective potential has the above form with $\eta = \eta(T)
= \sqrt{(T_c^2 - T^2)/6}$.\footnote{In this case, the change of the
tension is too mild. But, for example, the change of the tension for an
embedded domain wall after the core phase transition is significant due
to thermal effects.\cite{BWY}.} Here $T$ is the cosmic temperature and
$T_c$ is the critical temperature given by the coupling constants and
the zero temperature breaking scale. Thus, $\eta$ depends on the
temperature $T$, and hence on the cosmic time $t$,
$\eta=\eta(t)$.\footnote{Strictly speaking, the presence of cosmic
strings has effects on the finite temperature effective potential so
that $\eta$ can differ at points inside and outside strings. Here, we
assume that such backreaction effects are sufficiently small.} This
implies that the tension $\mu$ also depends on the cosmic time, $\mu =
\mu(t)$. As a more interesting example, we consider a real scalar field
$\chi$ oscillating around the origin with a mass and a quartic self
coupling, which may be the cold dark matter or the self-interacting cold
dark matter \cite{CDM}. We also assume that $\chi$ field has also a
coupling to the $\phi$ field of the form of $|\phi|^2 \chi^2$. That is,
the potential is given by
\beq
  V(\phi,\chi) = \frac{\lambda}{4}(|\phi|^2 - \chi^2)^2 
                + \frac12 m_{\chi}^2 \chi^2 + \frac{g}{4}\chi^4
\eeq
If a coupling constant $\lambda$ is small enough, the backreaction to
the oscillation of $\chi$ is negligible and $\eta$ in
Eq. (\ref{eq:potential}) is given by the root mean square of the
expectation value of $\chi$. Thus, for example, in case that the
oscillation of $\chi$ is dominated by the mass term, $\eta^2 \propto
a^{-3}$($a$: the scale factor), which is proportional to $t^{-3/2}$ in
the radiation domination and $t^{-2}$ in the matter domination. Hence,
the tension $\mu$ depends on the power of the cosmic time. Furthermore,
in the warped geometry, the expectation value $\eta$, that is, the
tension $\mu \simeq \eta^2$ depends on the position of the brane in the
bulk, which can depend on the cosmic time before the radion is fixed
completely. Thus, the case often takes place in which the tension $\eta$
of cosmic strings associated with the effective potential
Eq. (\ref{eq:potential}) changes with the cosmic time.

Then, we wonder whether the scaling solution also applies to cosmic
strings with time-dependent tension. In this paper we discuss this topic
in detail. In the next section, we derive the effective action for such
cosmic strings, which corresponds to the Nambu-Goto action for strings
with constant tension. In section III, we discuss the cosmological
evolution of such strings by considering the equation of motion in an
expanding universe. In the final section, we give our conclusions and
discussion.

\section{Effective action for cosmic strings with time-dependent tension}

Cosmic strings appear in the Abelian-Higgs model with a broken $U(1)$
symmetry, whose Lagrangian density is given by
\beq
  \CL = |D_{\mu}\phi|^2  
          - \frac14 F_{\mu\nu} F^{\mu\nu}
          - \frac{\lambda}{4}(|\phi|^2 - \eta^2)^2,
\eeq
where $\phi$ is a complex scalar field, $A_{\mu}$ are gauge fields,
$F_{\mu\nu}$ are field strengths, and $D^{\mu}\phi \equiv (\del^{\mu} -
i e A^{\mu})\phi$. The expectation value $\eta$ is now assumed to be
constant, and $e$ and $\lambda$ are coupling constants. Then, there
exists a static cylindrically-symmetric solution $(\phi_s, A^{\mu}_s)$
of the form \cite{Abrikosov,NO},
\bea
  \phi_s(r,\theta) &=& \eta e^{i n \theta} f(r), \non \\
  A^{\theta}_s(r,\theta) &=& \frac{n \alpha(r)}{er},
\eea
where $n$ is the winding number, and $f(r)$ and $\alpha(r)$ satisfy the
following equations,
\bea
 && \frac{d^2f}{dr^2} + \frac1r \frac{df}{dr} 
   - \frac{n^2f}{r^2}(\alpha-1)^2 
   - \frac{\lambda}{2} \eta^2 f(f^2-1) = 0,
 \non \\
 && \frac{d^2\alpha}{dr^2} - \frac1r \frac{d\alpha}{dr} 
     - 2e^2 \eta^2 f^2 (\alpha-1) = 0
\eea
with the boundary conditions $f(0)=\alpha(0)=0$ and
$f(\infty)=\alpha(\infty)=1$. Asymptotically $(r \rightarrow \infty)$,
$f(r)$ and $\alpha(r)$ can be approximated as
\bea
  f(r) &\sim& 1 
       - \CO \lmk \exp{\lmk -r/\delta_s \rmk} \rmk, \non \\
  \alpha(r) &\sim& 1 
       - \CO \lmk \sqrt{r \eta} \exp{\lmk -r/\delta_v \rmk} \rmk
\eea
for $\lambda/(2e^2) \lesssim 4$. Here $\delta_s \equiv
(\sqrt{\lambda}\eta)^{-1}$ and $\delta_v \equiv (\sqrt{2}e \eta)^{-1}$
are typical radii of cosmic string cores.

Now we consider the Abelian-Higgs model with the potential
Eq. (\ref{eq:potential}) in an expanding universe, whose action is given
by
\beq
  S = \int d^4 y \sqrt{-g} 
        \lkk |D_{\mu}\phi|^2  
          - \frac14 F_{\mu\nu} F^{\mu\nu}
          - V(\phi) \rkk.
 \label{eq:action}
\eeq
Here, note that the expectation value $\eta$ is now assumed to be
time-dependent, $\eta = \eta(t)$. The metric is taken to be that of the
spatially flat expanding universe,
\beq
  ds^2 = g_{\mu\nu} dx^{\mu} dx^{\nu}
       = dt^2 - a^2(t) d\vect x^2
       = a^2(\tau) (d\tau^2 - d\vect x^2)
\eeq
with $d\tau = dt / a(t)$ and $a(t)$ the cosmic scale factor.  The
worldsheet swept by a cosmic string can be characterized by two
parameters $\zeta^a$($a = 0, 1$) with
\beq
  x^{\mu} = x^{\mu}(\zeta^{a}).
\eeq
We take the timelike coordinate $\zeta^0$ to be the conformal time
$\tau$ and the spacelike coordinate $\zeta^1$ to be $l$, which
parametrizes the string at a fixed time. Then, the metric of the
spacetime can be written as
\bea
  ds^2 &=& g_{\mu\nu} dx^{\mu} dx^{\nu} \non \\
       &=& \gamma_{ab} d\zeta^{a} d\zeta^{b} ,
\eea
where $\gamma_{ab} \equiv g_{\mu\nu} x^{\mu}_{,a} x^{\nu}_{,b}$ and
$\gamma^{ab}$ is defined such that $\gamma_{ab} \gamma^{bc} =
\delta_{a}^{c}$. The comma represents the partial derivative. Since we
are interested in only the transverse motion of cosmic strings, without
generality, we can require the metric to satisfy
\beq
  \dot{\vect{x}} \cdot \vect{x}' = 0 \quad
  \Longleftrightarrow \quad \gamma_{01}=\gamma_{10}=0,
\eeq
where dots and primes represent derivatives with respect to conformal
time $\tau$ and the spacelike parameter $l$, respectively.

Now we derive an effective action for the above strings
\cite{effective,VS}. By introducing two spacelike vectors $n_{\mu}^A$,
$(A = 1, 2)$ with $g^{\mu\nu} n_{\mu}^A n_{\nu}^B = -\delta^{AB}$ and
$n_{\mu}^A x^{\mu}_{,a} = 0$, any point $y^{\mu}$ around the worldsheet
can be reparametrized as
\beq
  y^{\mu}(\xi) = x^{\mu}(\zeta) + \rho^a n^{\mu}_{A}(\zeta),
\eeq
where $\rho^A$ are two radial coordinates and $\xi^{\mu} =
(\zeta^a,\rho^A)$. The new coordinates $\xi^{\mu}$ are determined
uniquely if the point $y^{\mu}$ is inside the string's curvature
radius. Then, the string configuration can be approximated as
\bea
  \phi(y(\xi)) &=& \phi_s(r), \non \\
  A^{\mu}(y(\xi)) &=& n_{B}^{\mu} A_{s\,B}(r),
\eea
where $r^2 \equiv (\rho^1)^2 + (\rho^2)^2$ and $(\phi_s, A^{\mu}_s)$ are
a cylindrically-symmetric solution with $\eta =
\eta(\tau)$.\footnote{For a moving string, we use its Lorentz-boosted
version. Strictly speaking, time-dependent tension breaks Lorentz
invariance, which causes ambiguity for tension. But, the change of the
tension is not so rapid in our paper that the error associated with the
ambiguity is small and that we can safely use the tension with cosmic
conformal time $\tau$.} Here, we assume that the tension of a cosmic
string does not change so rapidly that its configuration can readjust
with the change of its tension, that is, $\delta_s, \delta_v \ll
\eta/(d\eta/dt)$. The Jacobian associated with the coordinate
transformation from $y^{\mu}$ to $\xi^{\mu}$ becomes $\sqrt{-\gamma}$
with $\gamma \equiv {\rm det} (\gamma_{ab})$ up to curvature terms
$\CO(\delta_s/R), \CO(\delta_v/R)$ with $R$ the typical curvature of the
string. On the other hand, integration of Eq. (\ref{eq:action}) over the
coordinates $\rho^A$ yields the time-dependent tension $\mu(t) \sim
\eta(t)^2$ up to curvature terms.\footnote{In the case that typical
curvature $R$ of the string is the horizon scale, which is expected in
the scaling regime, curvature terms are higher-order effects than that
of the variation of the tension. This is mainly because its typical time
scale of the variation is the Hubble scale when $\mu \propto t^{q}$ as
adopted later.} Thus, neglecting higher-order curvature terms, we obtain
the following effective action for a cosmic string with time-dependent
tension,
\beq
  S_{\rm eff} = - \int d^2 \zeta \sqrt{-\gamma} \,\mu(\tau).
\eeq
This is quite similar to the Nambu-Goto action with an additional factor
for the time-dependent tension. Note that $S_{\rm eff}$ is invariant
{\it neither} under the general coordinate transformation of $x^{\mu}$
{\it nor} under the reparametrization of $\zeta^{a}$. However, in the
cosmological situations, such an effective action can naturally appear
as a result of a cosmic string with time-dependent tension.

\section{Cosmological evolution of cosmic strings with time-dependent tension}

In this section, we will discuss the cosmological evolution of cosmic
strings with time-dependent tension. The Euler-Lagrange equation for the
effective action is given by
\beq
  \mu x^{\mu\,;a}_{,a} + \mu^{,a} x^{\mu}_{,a} 
   + \mu \Gamma^{\mu}_{\nu \sigma} \gamma^{ab} 
     x^{\sigma}_{,a} x^{\nu}_{,b} = 0,
\eeq
where the semicolon represents the covariant
derivative. $\Gamma^{\mu}_{\nu \sigma}$ is the four-dimensional
Christoffel symbol and is given by
\beq
  \Gamma^{\mu}_{\nu \sigma} = \frac12 g^{\mu \tau}
    (g_{\nu\tau,\sigma} + g_{\tau\sigma,\nu} - g_{\nu\sigma,\tau}).
\eeq
This equation can be rewritten as
\beq
   \frac{\mu}{\sqrt{-\gamma}} 
      \del_{a}(\sqrt{-\gamma} \gamma^{ab} x^{\mu}_{b})
   + \mu^{,a} x^{\mu}_{,a} 
   + \mu \Gamma^{\mu}_{\nu \sigma} \gamma^{ab} 
     x^{\sigma}_{,a} x^{\nu}_{,b} = 0.
\eeq

The $\mu = 0$ component of the equation of motion is given by
\beq
  \dot{\epsilon} + \frac{\dot{\mu}}{\mu} \epsilon 
   + 2 \frac{\dot{a}}{a} \epsilon \vect{x}^2 = 0,
\eeq
where $\epsilon \equiv
\sqrt{\vect{x}^{\prime\,2}/(1-\dot{\vect{x}}^2)}$. The $\mu = i$
component is given by
\beq
  \ddot{\vect{x}} 
    + 2 \frac{\dot{a}}{a} (1-\dot{\vect{x}}^2) \dot{\vect{x}}
    - \frac{1}{\epsilon} (\epsilon^{-1} \vect{x}')^{\prime} = 0.
\eeq

The energy $E$ of a cosmic string in an expanding universe can be
defined as
\beq
  E = a(\tau) \mu(\tau) \int dl \epsilon.
\eeq
Defining the energy density as $\rho \equiv E/V$ with $V$ some relevant
volume, the rate of change of $\rho$ is given by
\beq
  \dot{\rho} = -2 \frac{\dot{a}}{a} (1 + \la v^2 \ra) \rho,
\eeq
where $\la v^2 \ra$ is the average velocity squared of a cosmic string
defined as
\beq
  \la v^2 \ra \equiv \frac{\int dl \epsilon \dot{\vect{x}}^2}
                          {\int dl \epsilon}.
\eeq
Note that multiplying the rate equation of $\rho$ by $d\tau / dt$ gives
an equation with the same form but a derivative taken with respect to
the cosmic time $t$. Since it is more convenient to deal with $t$,
hereafter the dot represents the derivative with respect to the cosmic
time.

In order to investigate the evolution for such a string, we define a
characteristic scale $L$ of the string network as
\beq
  \rho_{\infty} = \frac{\mu(t)}{L^2},
  \label{eq:rhoinfty}
\eeq
where $\rho_{\infty}$ is the energy density of a string whose length is
larger than the horizon scale (called infinite strings). Furthermore, in
order to incorporate intercommutation effects,\footnote{Intercommutation
is a local process near the string core. Since $\delta_s, \delta_v \ll
\eta/(d\eta/dt)$, the time dependence of the tension has little effect
on the intercommutation.} we assume that the rate of energy transfer
from infinite strings to loops is given by
\beq \dot{\rho}_{\infty \rightarrow {\rm loops}} = c
  \frac{\rho_{\infty}}{L}, \eeq
where $c$ is a constant which parametrizes the efficiency of energy
transfer. Then, the rate equation for the energy density of infinite
strings becomes
\beq
  \dot{\rho}_{\infty} = 
    - 2 \frac{\dot{a}}{a} (1 + \la v^2 \ra) \rho_{\infty}
    - c \frac{\rho_{\infty}}{L}.
\eeq
Inserting Eq. (\ref{eq:rhoinfty}) into this equation yields
\beq
  -2 \frac{\dot{L}}{L} + \frac{\dot{\mu}}{\mu} =
    - 2 \frac{\dot{a}}{a} (1 + \la v^2 \ra) - \frac{c}{L}.
\eeq
Here we have assumed the constancy of $\la v^2 \ra$. In the scaling
regime, it seems reasonable just as the case with the constant
tension. But, strictly speaking, it must be checked by numerical
calculations, which will be done in near future.

In order to investigate the time development of $L$, we define $\gamma$
as $L = \gamma t$ and assume that $\mu \propto t^{q}$. Then, in a
radiation-dominated universe, the above equation can be recast into
\beq
  \frac{\dot{\gamma}}{\gamma} = - \frac{1}{2t}
     \lkk 1 - q - \la v^2 \ra - \frac{c}{\gamma} \rkk.
\eeq
Hence, the stable fixed point $\gamma_{\rm r}$ is given by
\beq
  \gamma_{\rm r} = \frac{c_{\rm r}}{1 - q - \la v^2_{\rm r} \ra}.
\eeq
In the same way, the stable fixed point in a matter-dominated universe
$\gamma_{\rm m}$ is given by
\beq
  \gamma_{\rm m} = \frac{3c_{\rm m}}{2 - 3q - 4\la v^2_{\rm m} \ra}.
\eeq 
Therefore, the characteristic scale $L$ of a string network also scales
with time for strings with time-dependent tension. That is, the number
of infinite strings per horizon volume is a constant irrespective of
time. But it should be noted that, due to the time dependence of the
tension, the ratio of the energy density of infinite strings to that of
the background universe is {\it not} a constant.

\section{Conclusions and Discussion}

In this paper, we have discussed the cosmological evolution of cosmic
strings with time-dependent tension. In all works about cosmic strings
which have been done thus far the tension of the strings has been
constant. However, strings with time-dependent tension often appear in
cosmological situations. In this paper, first we have derived the
effective action for such strings which, with an additional factor for
the time-dependent tension, is the Nambu-Goto action. Next, we gave the
equation of motion in an expanding universe and investigated the
evolution of such strings. We confirmed that, in the case where the
tension changes as the power $q$ of time, the string network obeys the
scaling solution---implying that the characteristic scale of the string
network grows with cosmic time. However, the ratio of the energy density
of infinite strings to that of the background universe is {\it not}
necessarily constant due to the time dependence of the
tension. Interestingly, the scaling may not be realized if $q$ is larger
than the critical value $q_r = 1 - \la v^2_{\rm r} \ra$ in a
radiation-dominated universe, or $q_m = 2(1 - 2\la v^2_{\rm m} \ra)/3$
in a matter-dominated one. But if this is the case, the change in the
tension is rapid enough that the effective action $S_{\rm eff}$ may not
be a good approximation. We will investigate this problem in the
future. Furthermore, numerical simulations are necessary to confirm our
results and constrain several parameters.

\section*{Acknowledgments}

M.Y.\ is grateful to Masahiro Kawasaki and Jun'ichi Yokoyama for useful
comments. M.Y.\ also thanks W. Kelly for correcting the English. M.Y.\
is supported in part by the project of the Research Institute of Aoyama
Gakuin University.

\end{document}